\documentclass[journal]{IEEEtran}
\usepackage{color}
\usepackage{multirow}
\usepackage{cite}
\usepackage{balance}
\usepackage{amsmath}
\usepackage{enumitem}
\setlist[itemize]{noitemsep, topsep=0pt}
\usepackage{amssymb}
\usepackage{textcase}
\usepackage[T1]{fontenc}
\usepackage[table,x11names]{xcolor}
\DeclareMathAlphabet{\mathpzc}{OT1}{pzc}{m}{it}
\usepackage{pdfsync}
\definecolor{light-gray}{gray}{0.88}
\usepackage[ruled,vlined,linesnumbered]{algorithm2e}
\usepackage{xcolor}
\ifCLASSINFOpdf
\usepackage[pdftex]{graphicx}

\DeclareGraphicsExtensions{.pdf, .jpeg, .png, .jpg, .gif}
\else
\usepackage[dvips]{graphicx}

\DeclareGraphicsExtensions{.eps}
\fi
\graphicspath{{figs/}} 
\usepackage{epstopdf}
\usepackage{booktabs}
\usepackage{tcolorbox}
\usepackage{amsmath}

\title{Risk-based Probabilistic Quantification of Power Distribution System Operational Resilience}

\begin{document}

\author{Shiva Poudel,~\IEEEmembership{Student Member,~IEEE,}
		Anamika Dubey,~\IEEEmembership{Member,~IEEE,} and Anjan Bose~\IEEEmembership{Life Fellow,~IEEE}% <-this % stops a space
		\thanks{S. Poudel, A. Dubey, and A. Bose  are with the School of Electrical Engineering and Computer Science, Washington State University, Pullman, WA, 99164 e-mail: shiva.poudel@wsu.edu, anamika.dubey@wsu.edu, bose@wsu.edu}% <-this % stops a space
	}

\maketitle
\vspace{-0.5 cm}
\begin{abstract}
    It is of growing concern to ensure the resilience in electricity infrastructure systems to extreme weather events with the help of appropriate hardening measures and new operational procedures. An effective mitigation strategy requires a quantitative metric for resilience that can not only model the impacts of the unseen catastrophic events for complex electric power distribution networks but also evaluate the potential improvements offered by different planning measures. In this paper, we propose probabilistic metrics to quantify the operational resilience of the electric power distribution systems to high-impact low-probability (HILP) events. Specifically, we define two risk-based measures: Value-at-Risk ($VaR_\alpha$) and Conditional Value-at-Risk ($CVaR_\alpha $) that measure resilience as the maximum loss of energy and conditional expectation of a loss of energy, respectively for the events beyond a prespecified risk threshold, $\alpha$. Next, we present a simulation-based framework to evaluate the proposed resilience metrics for different weather scenarios with the help of modified IEEE 37-bus and IEEE 123-bus system. The simulation approach is also extended to evaluate the impacts of different planning measures on the proposed resilience metrics.
\end{abstract}

\begin{IEEEkeywords}
    Distribution system resilience, Resilience metric, Value at Risk (VaR), Conditional Value at Risk (CVaR)
\end{IEEEkeywords}

\IEEEpeerreviewmaketitle

\section{Introduction}
\IEEEPARstart{T}{he} staggering cost of power system outages due to natural disasters combined with the impacts on personal safety and security from loss of critical services calls for an urgent need to ensure resilience in complex electric power networks \cite{reportirma,house2013economic}. The need for resilience is particularly critical for the aging power distribution grids, which are responsible for an estimated 90\% of the outages. This calls for a proactive disruption-management paradigm for power distribution systems beyond the classical reliability-oriented view that is driven by high-impact low-probability (HILP) events rather than persistent costs \cite{panteli2015grid}. Planning for resilience requires a metric that can not only quantify the impacts of potential HILP events on the grid but also help evaluate/compare different planning alternatives for their contributions towards improving the grid's operational resilience.

The operational resilience of power distribution systems is characterized as the system's ability to respond to and recover from an HILP event.  Traditionally, the performance of the power distribution system is measured using post-event reliability metrics such as SAIDI, SAIFI, MAIFI that provide an evidence-based performance indication on how well a specific distribution grid responded to the normal chance failure events/outages. 
Unlike routine outages, adequately anticipating and responding to HILP events is inherently difficult as they are rare. The impacts of HILP events cannot be properly quantified using reliability metrics calling for further considerations for beyond the classical reliability-oriented view \cite{panteli2015grid}. 
 
 In related literature, several methods have been proposed to quantify the resilience of the power grid. Resilience and robustness of the power system modeled as a complex network are studied in \cite{arab2015stochastic}; however, there exist a few indicators to compare the effectiveness of different proactive measures on power grid resilience. Others quantify resilience based on: 1) the availability of power resources to supply the interrupted loads after an extreme event \cite{bajpai2018novel}, 2) the ability of the system to supply the critical loads with reduced resources \cite{chanda2018quantifying}, and 3) social welfare indicating a satisfaction level depending on power and water availability for loads or customers \cite{najafi2018power}; however, they do not specifically model the probabilistic nature of HILP events. Furthermore, the proposed resilience metrics were mostly dimensionless quantities making it difficult to associate those to real-world consequences. Numerous optimization-based restoration methods have also been proposed to quantify the resilience based on the amount and duration of critical loads restored \cite{wang2015self, gao2016resilience, 8421055, lei2016mobile, xu2019resilience}. Some approaches emphasize on resilience driven, adaptive restoration strategies \cite{resende2011service, shekari2015analytical}. However, similar to \cite{chanda2016defining}, these methods are reactive, i.e. they focus on responding to an event rather than the quantification of resilience for possible future HILP events. Furthermore, they are specific to a given system and are scenario dependent and therefore, do not explicitly quantify expected system performance under unseen HILP events. Resilience metrics similar to reliability measures such as expected energy not served (EENS) and loss of load expectation (LOLE) have also been proposed \cite{nedic2006criticality, allan2013reliability}; however, these methods mostly provide a reliability-oriented view and do not specifically model the impacts of HILP events. 
Similarly, authors in \cite{mousavizadeh2018linear} proposed a two-stage stochastic framework for analyzing the resiliency of distribution networks during disasters. It is validated that the smart grid facilities such as microgrids and distributed energy resources improve the resilience and recovery of the grid. However, they do not specifically characterize the impacts of HILP events. Therefore, the metric cannot be generalized to assess system performance for possible future extreme events. A few authors developed resilience metrics by measuring ``reduced consequences from failures'' or ``deviation of system performance'' \cite{rose2007economic, refe}. However, they do not explicitly model the system behavior during and after an extreme event. A framework to measure and assess the resilience of a distribution system based on customer benefits is proposed in \cite{kwasinski2016quantitative}. The approach, however, is based on the concept of service availability in reliability theory and do not explicitly model the impacts of extreme events.

While a significant body of literature exists in defining and quantifying power systems resilience, no formal resilience metric is universally accepted. The existing metrics pose one or more limitations including: (1) they are post-event measures and hence do not provide any indication on the potential impacts of a future event; (2) they do not specifically measure the impacts of HILP events on system operations and performance loss; (3) they do not provide a generalized framework to evaluate the impacts of potential planning measures on improving system resilience. Along these lines, a few have suggested using risk-based characterization for resilience \cite{arghandeh2016definition}. For example, \cite{carlson2012resilience} proposed a method to assess the resilience of critical infrastructure systems for risk-based resource planning.  
A probabilistic approach to assess and evaluate the time-dependent resilience for electric power transmission systems is proposed \cite{panteli2017metrics}. Similarly, a framework to determine the system parameters that lead to low risks of power systems failure is proposed in \cite{cassidy2016risk}. However, there remains a critical gap in the existing literature on power grid resilience quantification. First, the existing literature does not include specific risk-based metrics and a framework to evaluate the same for power distribution systems. Second, there is little to no effort on developing a detailed resilience quantification framework that can help compare improvements in power distribution grid resilience due to alternate planning measures. 

With these considerations, we propose a framework to evaluate the resilience of power distribution systems using risk-based quantitative measures: Value-at-risk ($VaR_\alpha$) and conditional-value-at risk ($CVaR_\alpha$). Various optimization problem using $CVaR_\alpha$, mainly dealing with unit commitment and wind generation uncertainty have been studied in the past \cite{zheng2015stochastic, zhang2017conditional}. These metrics also have been extensively used in risk-averse financial planning to manage the impacts of low-probability high-risk financial investments. Similar considerations apply when managing the impacts of HILP events making these suitable metrics to quantify not only the potential impacts but also for comparative evaluation of the potential benefits offered by alternate planning investments.  
The specific contributions of this paper are summarized below:
\begin{enumerate}[noitemsep,topsep=0pt,leftmargin=*]
    \item {\em Resilience Metric:} Risk-based resilience metrics are defined to quantify the operational resilience of power distribution systems when impacted by extreme events. The proposed metrics (1) provide an indication of the \textit{potential impacts} of a \textit{future HILP event}; (2) measure the \textit{expected performance} of the system during an extreme event; (3) measure consequences of an event on power system operation and delivery (in terms of MWh not served); and (4) quantify the effectiveness of \textit{potential resilience enhancement strategies and investments} thereby, satisfying all four recommended criteria for a grid resilience metric as identified in several high-level roadmap studies related to critical infrastructure resilience (for ex. \cite{watson2014conceptual}). 
    \item {\em Comprehensive Simulation Framework for Resilience Quantification:} A framework based on Monte-Carlo simulation study is proposed to evaluate the impacts of HILP events on distribution system performance and to quantify the risks posed by such events on system's resilience. The proposed framework is extended to quantify the effects of potential disruption-management solutions that a utility can employ for improving system resilience. The resilience is quantified using the proposed metrics by simulating different levels of damages in power distribution systems and the effects of different recovery mechanisms.
\end{enumerate}

\section{Proposed Resilience Metrics: Definitions} 
\begin{figure}[t]
    \centering
    \vspace{-0.1cm}
    \includegraphics[width=0.45\textwidth]{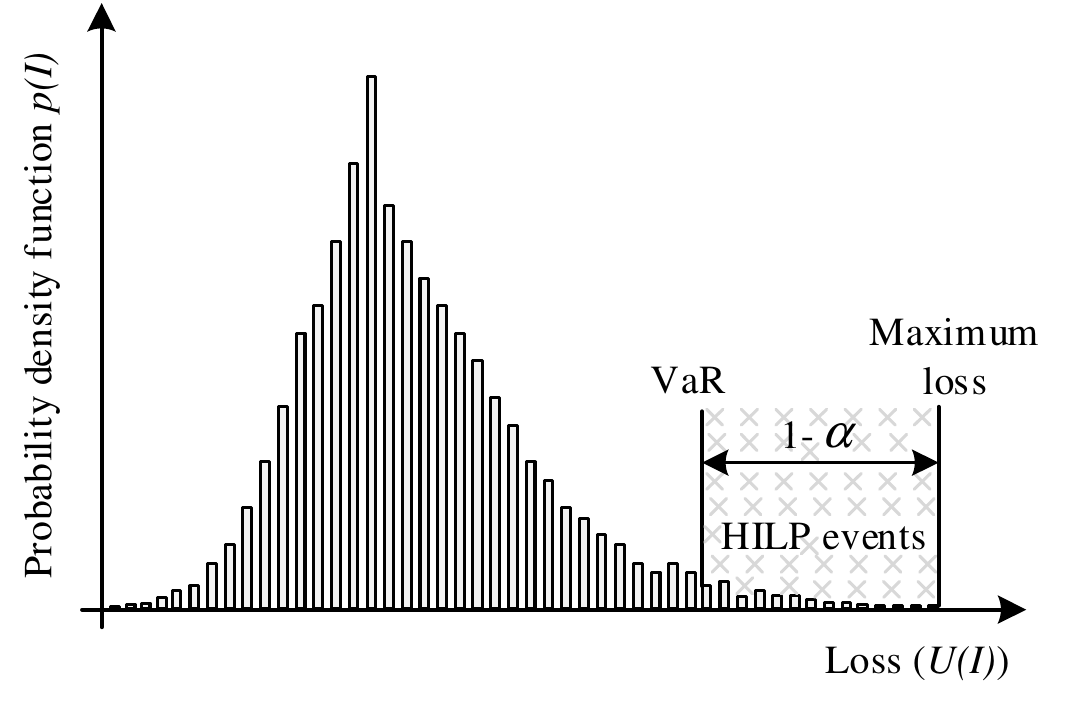}
    \vspace{-0.4cm}
    \caption{VaR and CVaR assessment for a probabilistic weather event. HILP events identified as the top $(1-\alpha)\%$ high impact disruptions.}
    \label{fig:1}
    \vspace{-0.1cm}
\end{figure}

\vspace{-0.0 cm} 
The proposed resilience metrics are motivated from risk-management literature that relates to quantifying the risks involved with a given financial investment. To this regard, $VaR_{\alpha}$ and $CVaR_{\alpha}$ are the two measures commonly used in risk-management literature to evaluate the impacts of low probability events that can potentially cause extreme losses for traders \cite{bardou2009computing}. $VaR_{\alpha}$ measures the maximum probable loss, while $CVaR_{\alpha}$ measures the expected shortfall due to the highest impact events  beyond a prespecified risk threshold, $\alpha$ (see Fig. \ref{fig:1}).
Since both metrics specifically quantify the extreme loses due to low probability events, they make suitable choices to quantify the operational resilience. 

In this section, we define several terms required to calculate the proposed resilience metrics for power distribution systems when subjected to HILP events. The system resilience is characterized based on the loss in system performance caused by probabilistic disruption events. To calculate this, we require a probabilistic model for the event and an approach to model its impact on system's performance loss function. 

\begin{figure}[t]
    \centering
    \includegraphics[width=0.45\textwidth]{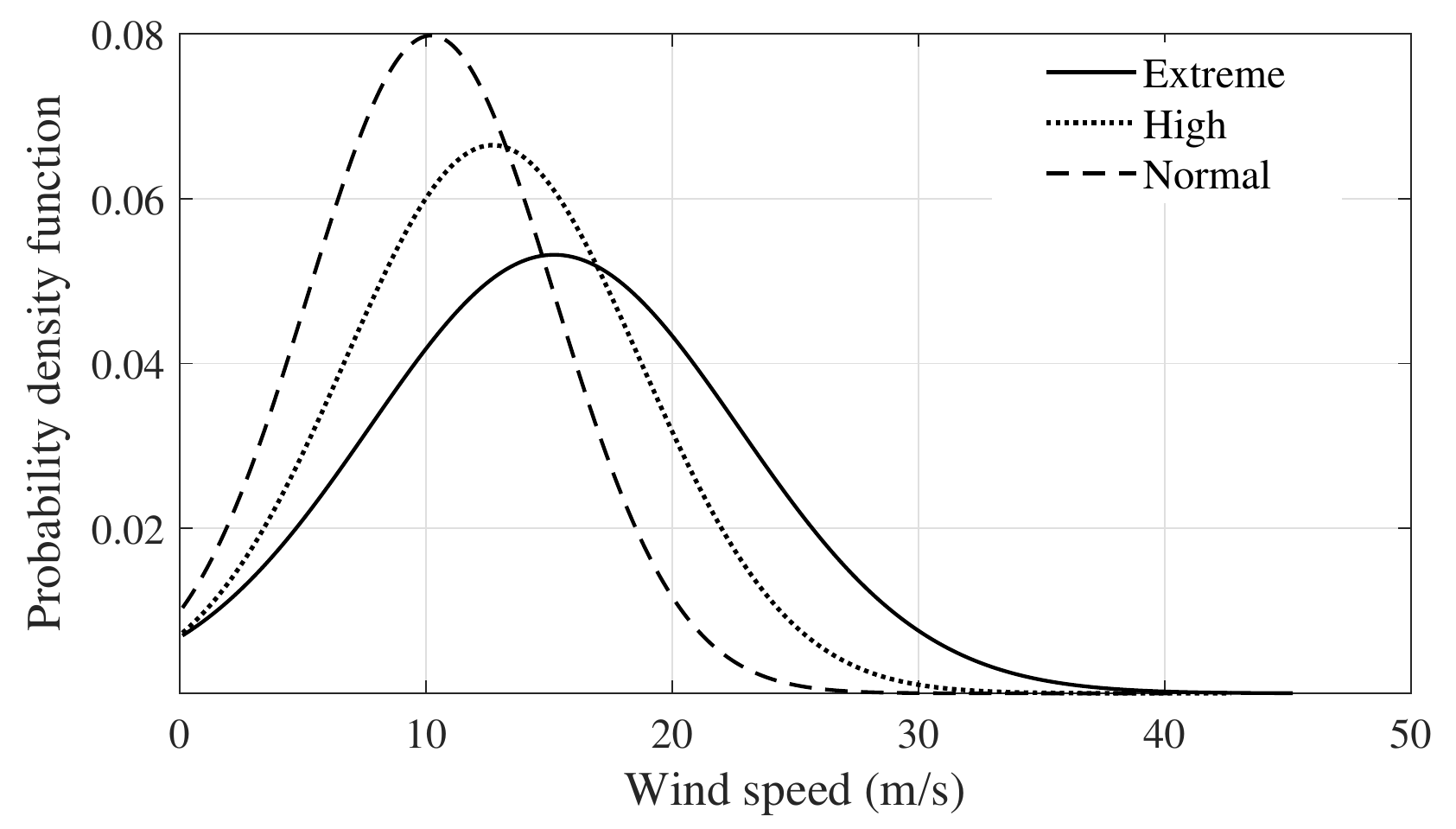}
    \vspace{-0.4cm}
    \caption{Regional wind profile}
    \label{fig:2}
    \vspace{-1.5cm}
\end{figure}
\subsection{Probabilistic Event and System Performance Loss} An event is characterized by two parameters - intensity of the event, $I$, modeled as a random variable; and the probability of its occurrence, $p(I)$. For example, the event of a windfall for three different regions (having extreme, high, and normal wind profiles) is characterized in Fig. \ref{fig:2}, where x-axis indicate wind speed (intensity of the event) and y-axis correspond to its probability. Fig. \ref{fig:2} represents the probability density function (PDF) of wind speed for a specific geographical region. 

The intensity of the event affects the failure probability of the system components (poles or lines) which in turn affects the system performance loss. A system performance curve also known as resilience curve is typically used to model the impacts of a specific disaster event on the distribution system performance \cite{panteli2017metrics}. The system performance loss, $U(I)$, when impacted by a random event $I$ is represented as a non-linear function of loss of load, $L(I)$, and total time taken to recover the system back to an acceptable level of performance, $t(I)$ as defined in (\ref{eq:u}). Simply, $U(I)$ is obtained by calculating the area under the system performance curve that measures the energy not served (MWh) in the aftermath of an event.
Note that $U(I)$ can be improved by proactive planning measures such as advanced restoration methods and hardening of distribution lines  \cite{mackenzie2016allocating}. These effects are modeled by appropriately representing the system performance curve as detailed in Section IV. 
\begin{equation}\label{eq:u}
    U(I) = f\left(L(I)\ t(I)\right), \ \textit{in MWh}
\end{equation}

\subsection{Risk-based Resilience Metrics} 
We define two risk-based resilience metrics: $VaR_{\alpha}$ and $CVaR_{\alpha}$. $VaR_{\alpha}$ calculates the maximum loss in resilience expected over a given time period for a specified degree of confidence, $\alpha$. . 
Simply, $VaR_{\alpha}$ refers to the lowest value, $\zeta$, such that with a probability $\alpha$, the loss does not exceed $\zeta$. $CVaR_{\alpha}$ measures the conditional expectation of the loss greater than those associated with $VaR_{\alpha}$.

To quantify system resilience, first, we calculate the system performance loss function, $U(I)$, by randomly sampling events from event pdf, $p(I)$. The pdf for $U(I)$ is obtained by associating the measured loss, $U(I)$, with the probability of observing such loss, i.e. $p(I)$ (see Fig. \ref{fig:1}). Let, $I$ represents a random variable corresponding to a weather event (e.g wind speed) having $p(I)$ probability of occurrence. Then, the probability that the system loss, $U(I)$, will not exceed a threshold $\zeta$ when impacted by a random event $I$ is given by (\ref{eq:2}). 

\begin{equation}\label{eq:2}
    \psi(\zeta)=\int_{U(I)\leq \zeta }^{} p(I) dI
\end{equation}

Here, $\psi$ is the cumulative distribution function (CDF) for the loss which determines the behavior of the system for a random event $I$. By definition, $VaR_{\alpha}$, with respect to a specified probability level $\alpha$ in $(0,1)$ is given by (3).
\begin{equation}\label{eq:var}
    VaR_{\alpha}= \text{min}\{\zeta \in \mathbb{R}:\psi(\zeta)\geq \alpha\}
\end{equation}

Next, $CVaR_{\alpha}$ metric is computed based on system performance loss caused by those probabilistic disruption events that cause the highest impacts \cite{vugrin2017resilience}. The metric $CVaR_{\alpha}$ defined in (\ref{eq:cvar}) measures the expected system loss (MWh) due to the top ($1-\alpha$)\% of highest impact events. It is worthwhile to mention that the events having \textit{losses greater than given level} and \textit{losses that do not achieve the prespecified level} are a complete group of events with probabilities $\alpha$ and $1-\alpha$ respectively. Thus, a $CVaR_{\alpha}$ metric defined in (\ref{eq:cvar}) measures the resilience of the system by calculating expected system performance loss due to the disruption events conditioned on the events being HILP (i.e. the tail of PDF for $U(I)$ (See Fig. \ref{fig:1})).
\begin{equation}\label{eq:cvar}
  CVaR_{\alpha}= (1-\alpha)^{-1}\int_{U(I)\geq VaR_{\alpha} }^{} U(I)\ p(I)\ dI.
\end{equation}

\section{Component-Level Impact Model}
Resilience metrics proposed in Section II-B require PDF for loss in system performance ($U(I)$) when impacted by a weather event. The probability of a weather event, $p(I)$, is obtained using the PDF for the specified weather event for a given region. For example, Fig. \ref{fig:2} shows PDF of regional wind speed for three different regions observing extreme, high, and normal wind speeds. 
The impact of a weather event on the distribution system is used to evaluate system performance loss function ($U(I)$). To do so, first, the probabilistic impact of a weather event, in this case, wind-related event, is modeled at the component-level. Next, a framework is developed to model the system-level impact of the weather event. 

\begin{figure}[t]
    \centering
    \includegraphics[width=0.5\textwidth]{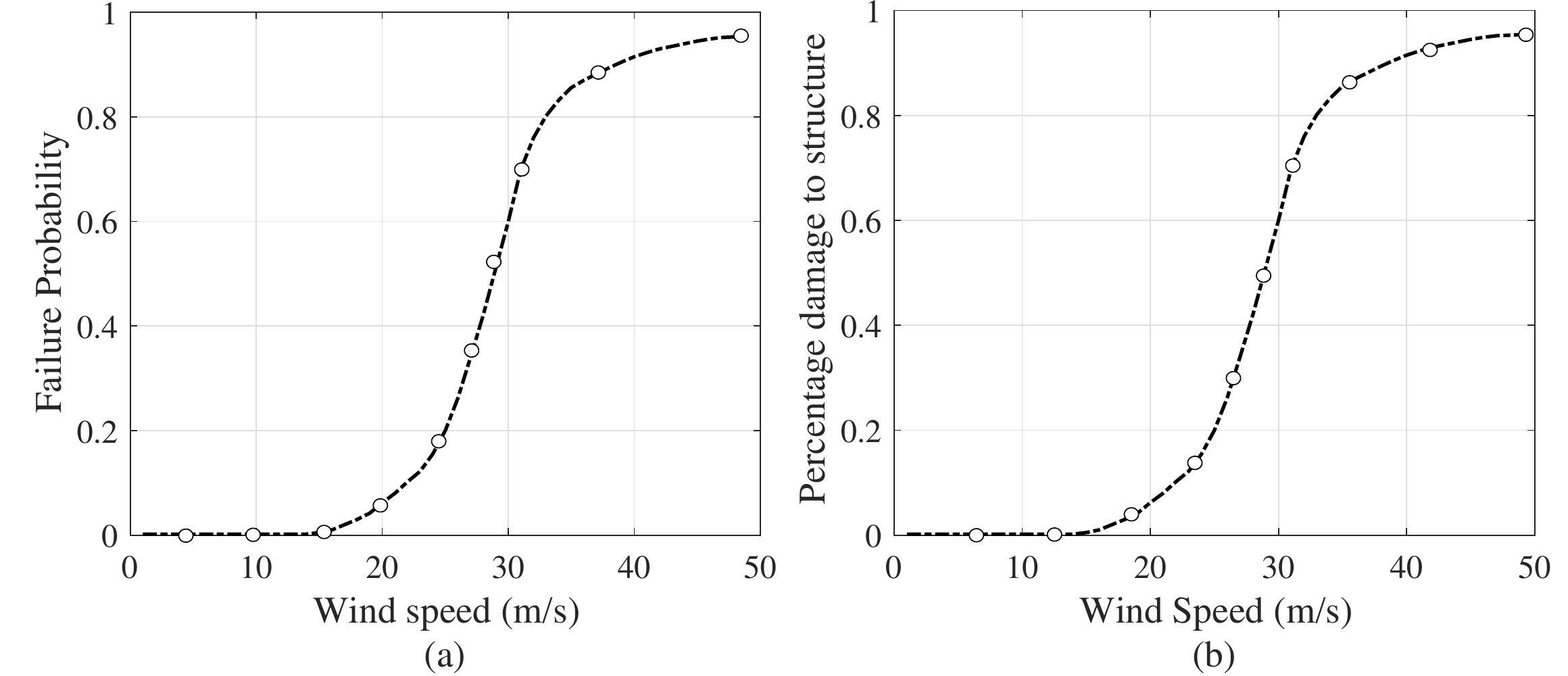}
    \vspace{-0.6cm}
    \caption{Component level impact assessment for an extreme event: (a) fragility curve for a wind profile \cite{panteli2017metrics} and (b) prototype curve fit models for percentage of equipment damaged as a function of wind speed \cite{powell1995real}.}
    \label{fig:3}
    \vspace{-0.6cm}
\end{figure}

\subsection{Probabilistic Component-level Fragility Curve} In related literature, component-level fragility curves have been used to model the impacts of hurricanes or other high-wind events on power system components \cite{panteli2017power}.
Specifically, a fragility function maps the probability of failure of distribution system components conditioned on the intensity of the hazard (e.g., a wind speed). An example of the fragility curve is shown in Fig. \ref{fig:3}a, that relates the failure probability of distribution system components to wind speed and is mathematically expressed as the following: 

  \[
    P_l(\omega)= 
    \begin{cases}
   P_l^n,& \text{if } \omega< \omega_{critical}\\
    P_l(\omega),& \text{if } \omega_{critical}<\omega< \omega_{collapse}\\
    1, & \text{if } \omega>\omega_{collapse}\\
    \end{cases}
    \]

\noindent where, $P_l(\omega)$ is the failure probability of a component as a function of wind speed, $\omega$; $P_l^n$ is the failure rate at normal weather condition; $\omega_{critical}$ is the wind speed at which the failure probability rapidly increases. The equipment has a negligible probability of survival at $\omega_{collapse}$.

The fragility curve can be generated using empirical data gathered during disaster events. For example, in \cite{powell1995real}, prototype curve-fit models for percentage damage in distribution system components are constructed as a function of maximum sustained wind speed using the recorded data on observed damage within a given region (see Fig. \ref{fig:3}b). The observed damage from the past wind storms can yield estimates of damage for a given wind profile.
These damage estimates can be coupled with infrastructure database and geographical information system to construct an equivalent fragility curve for the component. In this paper, the values of the fragility curve are selected randomly for simulation purposes. However, empirical data, if available, can be used to adjust the parameters. 

\vspace{-0.4cm}
\subsection{Generate Component Damage Scenarios}
The event-dependent failure probability of the distribution system components, $P_l(\omega)$, is used to identify the operational states of individual distribution system components. Corresponding to each component, $c$, a uniformly distributed random number, $r_k \sim \mathcal{U} (0,1)$, is generated. These random numbers are compared with the event-dependent failure probability $P_l(\omega)$ for the given wind speed, $\omega$, to obtain the operational status of the component, $F_l^c(\omega)$, using (\ref{eq:5}).
\begin{equation}\label{eq:5}
    F_l^c(\omega)= 
    \begin{cases}
    0,& \text{if } P_l(\omega)<r_k\\
    1,& \text{if } P_l(\omega)>r_k
    \end{cases}
\end{equation}
where, $F_l^c$ is the failure function of $c^{th}$ component where $F_l^c(\omega)=1$ implies failure and vice-versa. The wind-affected operational states for all distribution system components is obtained. The component-level damage is used to calculate system-level impacts and consequently the system loss function as detailed in the following section.  

\section{System-Level Impact Model} 
This section details the approach to model the impact of a weather event on the distribution system.
\vspace{-0.4cm}
\subsection{System Performance Curve/Resilience Curve}
\begin{figure}[t]
\vspace{-0.0cm}
    \centering
    \includegraphics[width=0.5\textwidth]{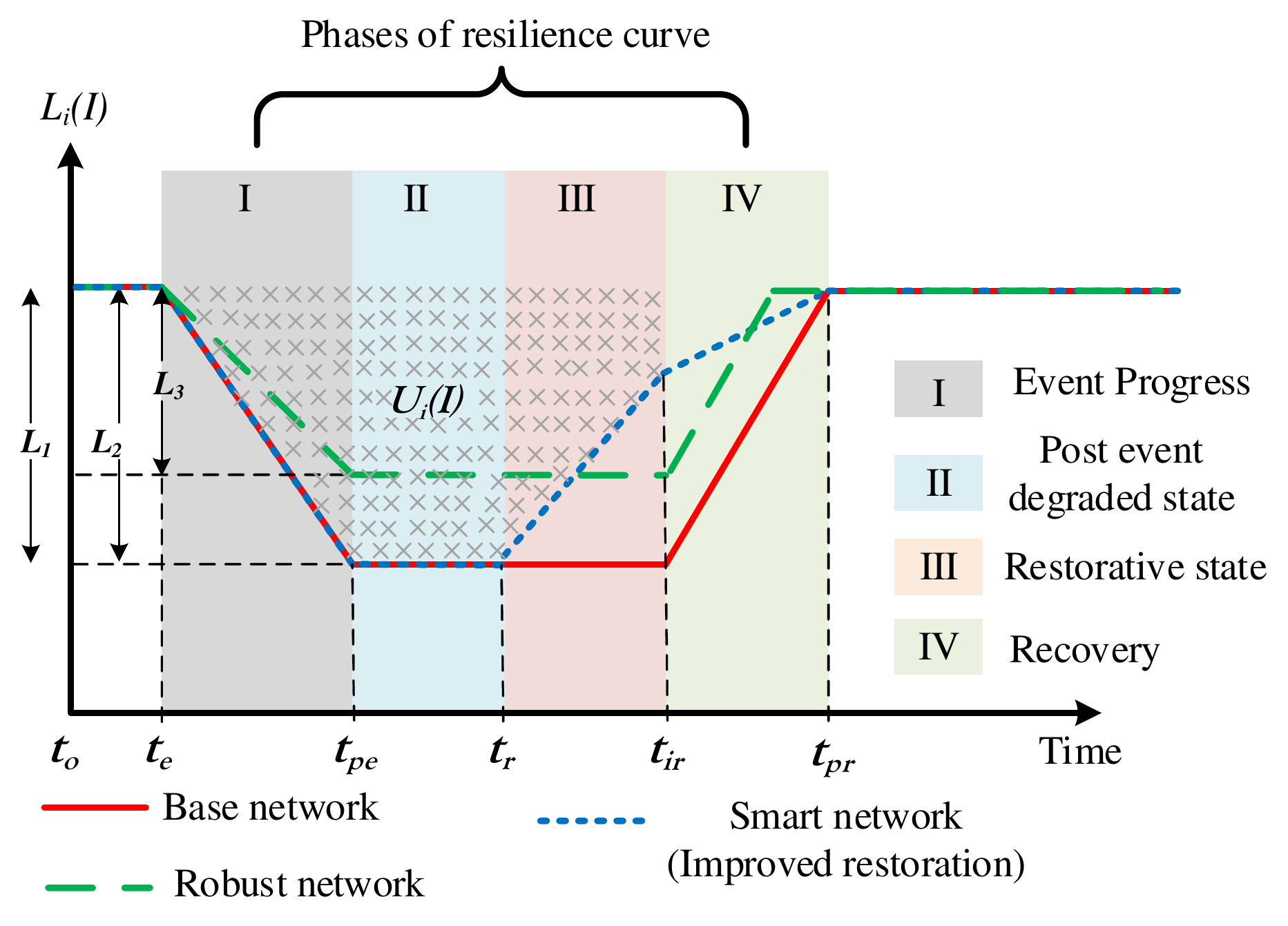}
    \vspace{-0.4cm}
    \caption{Approximated resilience curve for an event. The different colored lines correspond to effects of proactive planning: (1) Base network - does not include any proactive planning measure for resilience improvement; (2) Smart network - includes DGs to support intentional islanding during Phase III; (3) Robust network - includes hardening of the distribution lines lowering the probability of component damage.}
    \label{fig:4}
    \vspace{-0.5 cm}
\end{figure}

A simplified resilience curve demonstrating different phases in which the power distribution system resides in the aftermath of an extreme event and their time progression is shown in Fig. \ref{fig:4} \cite{panteli2017metrics}. These include event progress, post-event degraded state, restorative state, and recovery. The different phases as indicated in Fig. \ref{fig:4} are as following:
\begin{enumerate}[noitemsep,topsep=0pt,leftmargin=*]
    \item Phase I: Event progress $(t\in[t_e, t_{pe}])$, duration of event.
    \item Phase II: Post-event degraded state $(t\in[t_{pe}, t_{r}])$ following the end of event progress and before any restoration begins. Damage assessment is performed in this phase. 
    \item Phase III: Restorative state $(t\in[t_r, t_{ir}])$, for automated restoration prior to any infrastructure recovery.
    \item Phase IV: Infrastructure recovery stage $(t\in[t_{ir}, t_{pr}])$, system returns to original state prior to disaster.
\end{enumerate}

Since the focus is on operational resilience, infrastructure recovery phase is not included in the simulation framework for metric calculation. That is, system performance between $\{t_e, t_{ir}\}$ as represented in Phases I-III is used to characterize the operational resilience metric. However, infrastructure recovery can be easily included in the framework if enough information regarding the overall repair process is available. The information required for including infrastructure recovery stage includes the dispatch of crew members, the available resources for system repair, and the overall system recovery and repair process. The only change that is required would be the addition of area under the curve between time intervals $t_{ir}$ and $t_{pr}$ when evaluating the system performance loss for a particular event.

\subsection{Quantify System Performance Loss, $U_i(I)$}
System performance loss, $U_i(I)$, measures energy not served (MWh) in the aftermath of an event. It is a multidimensional concept that requires both system loss ($L_i$) and transition time between the phases/states. Therefore, the operational resilience is greatly influenced by the response of system during Phases I--III of the resilience curve. 

\subsubsection{Initial System Loss, Phase I ($t_{pe}-t_e$)}
The initial impact depends on the type of disruption. For example, when the system is hit by an earthquake, there is a sharp decrease in resilience. On the other hand, a wind-related event may take a longer time to degrade the system as it progresses geographically. In the state of event progress, the service to loads is interrupted and hence the loss function increases. If the time and location of the extreme event can be predicted accurately, preventive actions can be applied before the event hits the system and help decrease $L_i(I)$. Also, added redundancy can help reduce the system performance loss and/or reduce the slope of resilience degradation (See Phase-I of Fig. \ref{fig:4}).

\subsubsection{Damage Assessment, Phase II $(t_r-t_{pe})$}
Following the event, the system enters the post-event degraded stage ($t_{pe}$) where the resilience of the system is significantly compromised. At this stage, the system prepares for the restoration where damage assessment is done and the available resources and a healthy portion of network are identified with the help of utility crew members, collective information from smart meters, geographical information system. 
Damage assessment is crucial for enhancing resilience as no restoration actions can be initiated unless the systems' degraded state is known. The system remains in the post-degraded state until the damage assessment is complete which is represented by Phase-II in Fig. \ref{fig:4}. Note that the duration for damage assessment can be improved with the help of fault identification algorithms, optimized crew dispatch, outage report from customers, and aerial survey after any extreme condition. %\cite{epri}. 

In general, the time for damage assessment can be calculated using historical information on prior disasters. In this paper, for normal weather conditions, time for damage assessment, $DA^{time}_n$, is assumed to be 2 hours; $DA^{time}$ for other weather conditions is determined using (6). 
\begin{equation}
    DA^{time}= 
    \begin{cases}
   DA^{time}_{n},& \text{if } \omega \leq 20\ \text{m/s}\\
    n_1 \times DA^{time}_{n},& \text{if } 20\ \text{m/s}<\omega\leq 40\ \text{m/s}\\
    n_2 \times DA^{time}_{n}, ,& \text{if } \omega>40\ \text{m/s}\\
    \end{cases}
\end{equation}
where, $n_1 \sim \mathcal{U}_c (3, 4)$ and $n_2 \sim \mathcal{U}_c (5, 6)$ are random numbers generated within the pre-specified range. However, the accuracy of the simulation output can be improved using accurate values of $DA^{time}$ provided by the system operator for different intensities of weather event.

\subsubsection{Restoration and Active Islanding Scheme, Phase III $(t_{ir}-t_{r})$}
Following the actions in post-event degraded state, the system enters the restorative state where a proper restoration strategy is determined and disconnected loads are gradually restored with help of feeder reconfiguration, distributed generation (DGs) and other smart actions. It is a step-wise process where several switching actions are performed for the reconfiguration of the network. The presence of remote-controlled switches (RCS) and automated intentional islanding procedures using utility-owned DGs can help restore the system's critical loads prior to infrastructure recovery. This can help improve distribution system performance during the restorative state before the infrastructure recovery is initiated (i.e., from $t_r$ to $t_{ir}$). This is represented by Phase-III in the resilience curve as shown in Fig. \ref{fig:4}. It is understood that the slope of a restorative state depends on the level of impact, size of DGs available for active islanding, and the number of switching schemes required for forming the DG supplied islands. 

\begin{figure*}[t]
\vspace{-0.0 cm}
\centering
    \includegraphics[width=1.0\textwidth]{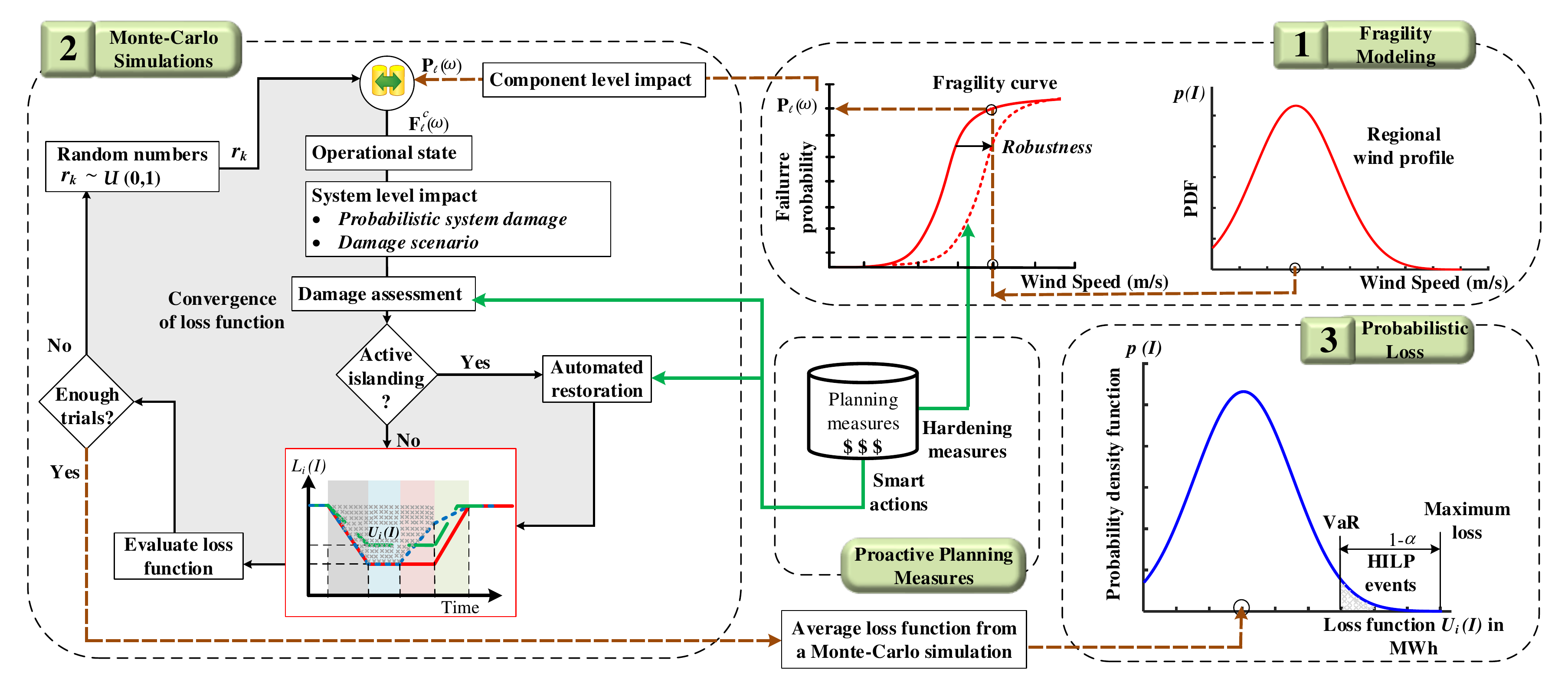}
    \vspace{-0.9 cm}
    \caption{From fragility modeling to probabilistic loss for resilience metric characterization. The steps for computing resilience metrics as outlined in Algorithm 1 are detailed here.}
    \label{fig:5}
    \vspace{-0.4 cm}
\end{figure*}

\subsubsection{Calculate System Performance Loss}
Based on the aforementioned discussion, the system performance loss, $U_i(I)$ when impacted by an event $I$ is given by the area under the resilience curve from $t_e$ to $t_{ir}$. For example, for a smart network, the performance loss is represented by the shaded region in Fig. \ref{fig:4}. The performance loss for three cases in Fig. \ref{fig:4} can be evaluated using (\ref{eq0}a), (\ref{eq0}b), and (\ref{eq0}c) respectively.
\begin{subequations} \label{eq0}
\begin{align}
U_1(I)&=\frac{1}{2}\big[(t_{ir}-t_{pe})+(t_{ir}-t_e)\big]L_1\\
U_2(I)&=\frac{1}{2}(2t_r-t_e-t_{pe})L_2+\frac{1}{2}(2L_2-P_{res})(t_{ir}-t_r) \\
U_3(I)&=\frac{1}{2}\big[(t_{ir}-t_{pe})+(t_{ir}-t_e)\big]L_3
\end{align}
\end{subequations}
where, $i$ is the notation for particular scenario: ($i=1$) base network, ($i=2$) smart network, ($i=3$) robust network; $P_{res}$ is the amount of load restored by DGs for the restorative state. It should be noted that the time from $t_{r}$ to $t_{ir}$ is equal to the time taken for performing the switching actions for restoration based on active islanding scheme. It is assumed that the switching scheme is sequential and the total switching time ($T_s$) is the sum of switching times (\ref{sw}). 
\begin{equation}\label{sw}
    t_{ir}-t_r=T_s=\sum_{m\in N_{sw}}t_s^m
\end{equation}
where, $t_s^m$ is the switching time for switch $m$; $N_{sw}$ is the set of switches involved in switching actions. The time to operate a switch pair depends upon switch type. We assume mean time to operate a manual switch and RCS to be 30 minutes and 20 seconds, respectively.

\begin{algorithm}[b]
    \label{algo}
    \textbf{Given}: {Weather data, Distribution system model}\\
    \colorbox{light-gray}{\textbf{Step I: Fragility Modeling}}\\
    Obtain PDF $p(I)$ of the wind-speed profile for a given geographical region using weather data \\
    \For {{each distribution lines}}{
    Generate fragility curves\\
    Obtain component failure probabilities $\textbf{P}_l(\omega)$
    }
    \colorbox{light-gray}{\textbf{Step II: Monte-Carlo Simulation}}\\
    \For {\text{each event in $I$}}{
    Component level impact$\to$ System level impact\\
    Evaluate system loss for given event $U_i(I)$\\
    \If{enough trials,}{
    Evaluate average loss function\\
    }
    \Else{Go to step 9}{
    }
    }
    \colorbox{light-gray}{\textbf{Step III: Probabilistic loss}}\\
    Compute risk-based resilience metrics\\
    \textbf{Output:} $VaR_\alpha,\  CVaR_\alpha$\\
    \caption{Probabilistic loss for a given HILP event}
\end{algorithm}

\subsection{Model the Impacts of Proactive Planning}
Several proactive planning measures can be applied to enhancing the distribution system resilience. From an operational standpoint, a decision-maker can improve resilience by allocating resources to lessen the average impact ($L_i$), decrease the damage assessment time ($t_{r}$-$t_{pe}$) to quickly enter the restorative state and/or apply advanced restoration to decrease impact in restorative state ($t_{ir}$-$t_{r}$). Two specific proactive planning measures and approach to model their impact on resilience curve are discussed in this section.
\subsubsection{Robust Network--Hardening/Under-grounding}
Hardening the distribution lines, although expensive, is one of the most effective methods to protect the system against extreme wind storms. The hardening solutions mainly boost the infrastructure resilience thus reducing the initial system loss, $L_i(I)$, as the event strikes and progresses. Several hardening strategies include overhead structure reinforcement, vegetation management, and undergrounding of distribution lines. By hardening, components are made robust against the extreme wind profile by reducing the probability of wind-induced damages. In essence, hardening modifies the fragility curve for the distribution system components (see Fig. 5). With this planning measure, a decision-maker can expect to have reduced loss, $L_i$ during the state of event progress as shown by the green curve in Fig. \ref{fig:4}. 
\subsubsection{Smart Network--Improved Response and Automated Restoration}
This is one of the smart solutions that aim at enhancing the operational resilience by quickly restoring the distribution system. Distribution circuit equipped with enough smart meters and RCS allows for advanced automation capabilities. In the presence of grid-forming DGs and RCSs, intentional islands can be formed to supply critical loads prior to infrastructure recovery as proposed in authors prior work \cite{8421055}. Besides, adequate situational awareness tools can enable effective and timely decision making for damage assessment and commencing appropriate actions for restoring the system's critical loads. This planning measure cannot help during the phase of event progress and is not capable of reducing the initial impact. However, it assists the system operator in quick damage assessment and smart restoration of the system that helps achieve the post-disturbance resilience by reducing the duration of the degraded state. This effect is characterized by  the blue curve in Fig. \ref{fig:4}, where area under the Phase III $(t_r,t_{ir})$ decreases due to restoration action. 

\section{Monte-Carlo Simulation to Characterize Distribution System's Operational Resilience}
In this section, we detail the process for computing the resilience metric by evaluating the probabilistic system damage. 
\vspace{-0.5 cm}
\subsection{Probabilistic Function for System Performance Loss}

In the proposed approach, the component level fragility curves are used to generate system loss function. The overall procedure introduced in Section III and IV is summarized in Algorithm 1 and is detailed in Fig. \ref{fig:5}. The framework requires weather data and detailed system model as input and it outputs the resilience metrics ($VaR_\alpha$ \& $CVaR_\alpha$). Since weather intensity and its impact on equipment are not deterministic, a Monte-Carlo simulation method, largely used in bulk-system reliability calculation \cite{ li2013reliability, allan1988monte,liang1997distribution}, is employed to evaluate the probabilistic impacts of a weather event on the distribution system. Although computationally intensive, Monte Carlo simulations are suitable for the analysis and planning purposes of complex systems as evident in related prominent literature in this domain \cite{kaygusuz2014monte, zhang2017conditional, paul2018resilient}. Furthermore, since the approach does not include any assumptions regarding the probabilistic events, as often employed in related analytical methods to reduce the complexity of probabilistic computations, the approach is generalizable and relatively more accurate. In this work, we aim at evaluating the resilience of a distribution system for future extreme events that are rare. Monte Carlo method enables simulation of such rare events while including the low probability of observing those to provide a realistic assessment of the risks associated with such HILP events. The overall process to characterize the distribution system's operational resilience is discussed below:
\subsubsection{Fragility Modeling}
The fragility curve for each distribution component, similar to one shown in Fig. 3, is generated for a given weather event (in this case wind-related events). Note that fragility curves can be generated empirically, experimentally or analytically using expert judgments as detailed in Section III-A. The PDF for the concerning weather event for the specified region is obtained using the meteorological data collected using weather sensors; in this case probabilistic wind-speed data (see Fig. 2). By mapping the regional wind speed PDF (Fig. \ref{fig:2}) to the component fragility curves (Fig. \ref{fig:3}), the components' failure probabilities ($P_l(\omega)$) are obtained. Recall that one of the proactive planning measures is hardening of the distribution lines. Hardening of the distribution lines will shift the fragility curve to the right thus making components more robust to higher intensities of weather event (see Fig. \ref{fig:5}). Thus, when hardening is available, the failure probability of the corresponding lines is changed.

\begin{figure*}[t]
    \centering
    \includegraphics[width=0.99\textwidth]{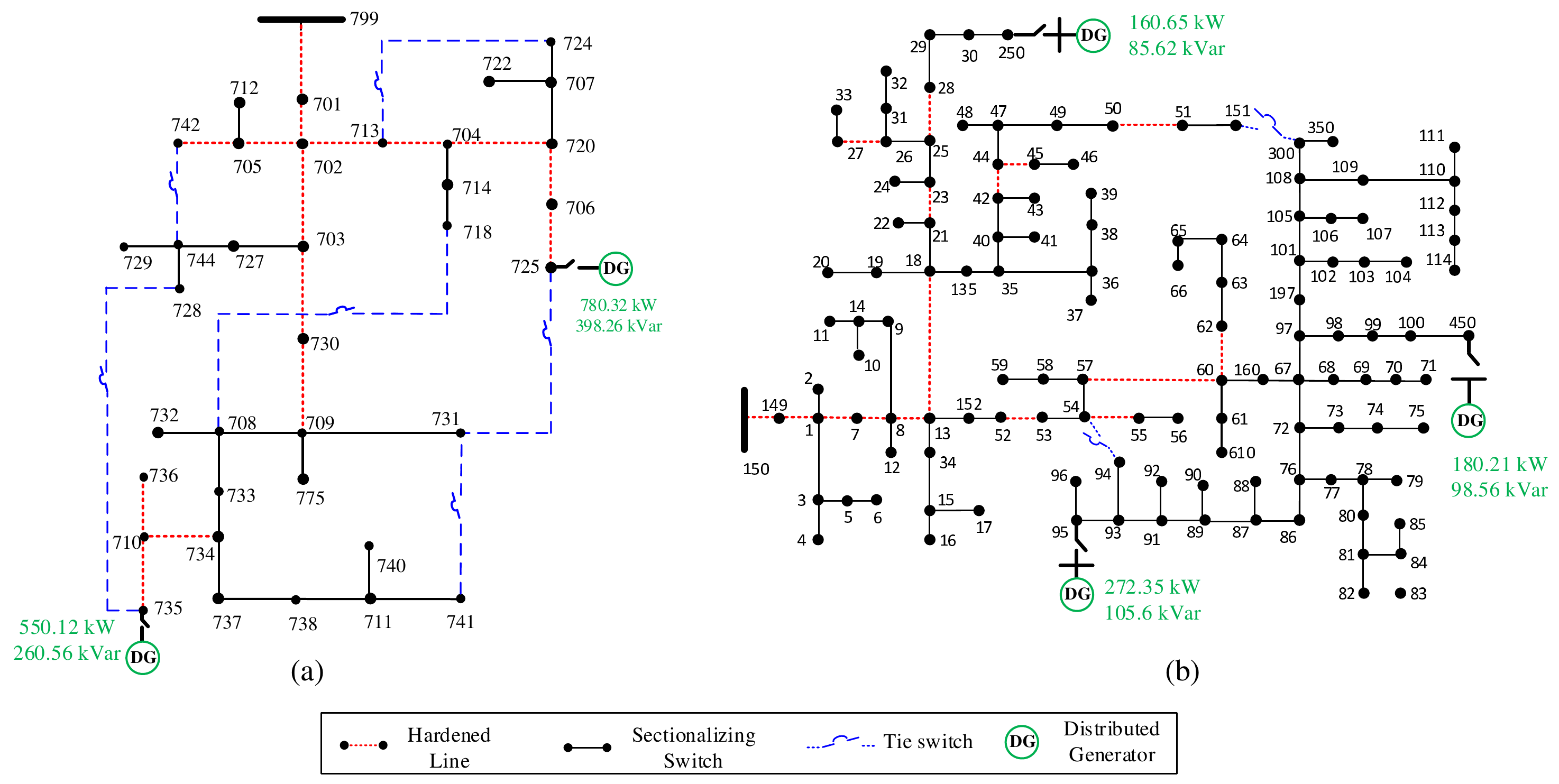}
    \vspace{-0.2cm}
    \caption{Modified test cases with DGs and tie switches: (a) IEEE 37-bus, (b) IEEE 123-bus }
    \label{fig:6}
    \vspace{-0.6cm}
\end{figure*}

\subsubsection{Monte-Carlo Simulations}
Monte-Carlo simulations are used to generate damage scenarios for particular wind speed. At each trial, the component-level failure probability, $P_l(\omega)$, is used to determine the operational state of the particular component, $F_l^c(\omega)$ using (5). Following a similar process, operational states of all distribution system components are obtained representing the damage scenario for a particular Monte-Carlo trial. The generated damage scenario is then used to evaluate the system loss function by generating the resilience curve. The initial loss function is obtained using distribution system simulator, OpenDSS. The effect of a smart network (improved response and automated restoration) is taken into consideration when evaluating the loss function if such measures are available. To do so, the restoration scenarios are embedded using the dedicated algorithms developed in authors prior work \cite{8421055}. Specifically, the restoration problem is formulated as a mixed-integer linear program (MILP) to maximize the load restored with the help of all available feeder and DGs with intentional island formation \cite{8421055}. 
After implementing the smart actions, Phases I, II, and III of the resilience curve are analyzed and loss function is calculated using (7). Several trials of Monte-Carlo simulations are done to obtain statistically representative results (See Fig. \ref{fig:5}). The process is repeated for multiple wind speeds. It should be noted that if smart strategies are not available, it is assumed that the system is restored only after the infrastructures are recovered i.e. in Phase IV.

\subsubsection{Probabilistic Loss}
After a sufficient number of Monte-Carlo simulations, average performance loss (in MWh) for each sample of wind speed is obtained. The average loss is then mapped onto the PDF for weather-event (wind speed) to get a probabilistic representation for system performance loss when subjected to a given weather event. At the end of the Monte-Carlo simulations for all sampled intensities of the weather event, i.e. wind speed, the PDF for loss function is obtained as shown in Fig. \ref{fig:5}. In this paper, loss in energy (MWh) is used to measure the operational resilience of distribution system subject to an extreme wind storm. The framework can be applied to derive similar probabilistic loss for other indices, for eg., loss of load, loss of critical load, loss of life, etc.

\subsection{Compute Risk-based Resilience Metric ($CVaR$ and $VaR$)}
The PDF representing for probabilistic loss function, $U_i(I)$, is used to compute $VaR_\alpha$ and $CVaR_\alpha$ metric using (\ref{eq:var}) and (\ref{eq:cvar}). For a given confidence level $\alpha$, $CVaR_\alpha$ measures the conditional expectation of observing a system loss (energy loss in MWh) due to $(1-\alpha)\%$ of highest impact events. The expected loss as evaluated from $CVaR_\alpha$ quantifies expected loss in energy in MWh due to extreme wind-related events.

 Note that the important distinction and novelty of the proposed metric and simulation approach lies in its ability to specifically model and quantify the effects of HILP events. Therefore, although $CVaR_{\alpha}$ quantifies MWh not served, drawing a close analogy with EENS -- a well-known reliability index used for bulk-grid, the distinction lies in its ability to specifically characterize the extreme events, unlike EENS that quantifies unserved energy for expected contingencies. \\ \\

\textit{Discussion on Applicability of the Proposed Simulation Framework to Realistic Distribution System Models:}
The proposed simulation framework can easily include any future distribution system configurations such as those with battery storage systems, photovoltaic generation (PVs), electric vehicles (EVs), etc. and also detailed feeder models including service transformers, secondary feeder, and individual customers loads. The ability to generalize the simulation model emanates from the use of OpenDSS to simulate distribution systems operations. OpenDSS is a distribution system simulator that is capable of analyzing a detailed electric power distribution grid with distributed generation resources and new load models such as EVs \cite{opendss}. Similar considerations apply when incorporating detailed protection systems model/configuration including mid-line reclosers, fuses, sectionalizers, etc. which may help better isolate the faulted sections and enable restoring additional loads on healthy but de-energized feeder sections. In this study, we employ a simplified protection system model for simulation studies. Essentially, it is assumed that each faulted/damaged line can be isolated with the help of switches, and the healthy but de-energized part of the feeder can then be restored using available feeders and DGs. The simulation framework, however, is applicable to any complex protection systems model that can be simulated using OpenDSS.

\section{Results and Discussions}
IEEE 37-bus and IEEE 123-bus test systems are used to demonstrate the proposed framework. The Monte Carlo simulations are carried out on a PC with 3.4 GHz CPU and 16 GB RAM. The restoration problem for each scenario is formulated as an MILP that can be solved using any off-the-shelf MILP solvers. We have used MATLAB R2016a to build the MILP model that is linked to the CPLEX 12.6 solver. On an average for a scenario, it takes about 2 seconds to solve the restoration problem for IEEE 123-bus and less than a second for IEEE 37-bus test case. The following assumptions are made for simulation purposes:
\begin{enumerate}[noitemsep,topsep=0pt,leftmargin=*]
    \item For the base case, we use same fragility curves for all distribution components assuming it spans a small geographic area and observes a homogeneous weather condition.
    \item It is assumed that no recovery/restoration is performed during the event progress ($t_o$ to $t_e$) and all the components are online before the event. 
    \item To model proactive disruption-management solutions for smart network, two tie switches, 54-94 and 151-300, are added to simulate different restoration scenarios and line 93-94 is upgraded to a three-phase line in IEEE 123-bus test case. Similarly, for IEEE 37-bus test case, six tie switches are added for creating restoration scenarios. In addition to new switches, utility-owned DGs with grid-forming inverters are used to assist with the restoration process via intentional islanding in both test cases (see Fig. \ref{fig:6}).
    \item To model proactive disruption-management solutions for robust network, 15 overhead lines are randomly selected and hardened for IEEE 37-bus whereas for IEEE 123-bus 16 different lines are hardened (see Fig. \ref{fig:6}). This shifts the fragility curve for the hardened lines to the right (see Fig. \ref{fig:5}). 
    \item With high likelihood, a suitable protection system for dynamically formed islands supplied by DG will not be available due to damages and incorrect protection settings. Thus, it is assumed that the existing protection system is disabled when implementing advanced restoration decisions by forming self-sustained islands using distributed generators and tie switches.

\end{enumerate}

 \begin{figure}[t]
    \centering
    \includegraphics[width=0.49\textwidth]{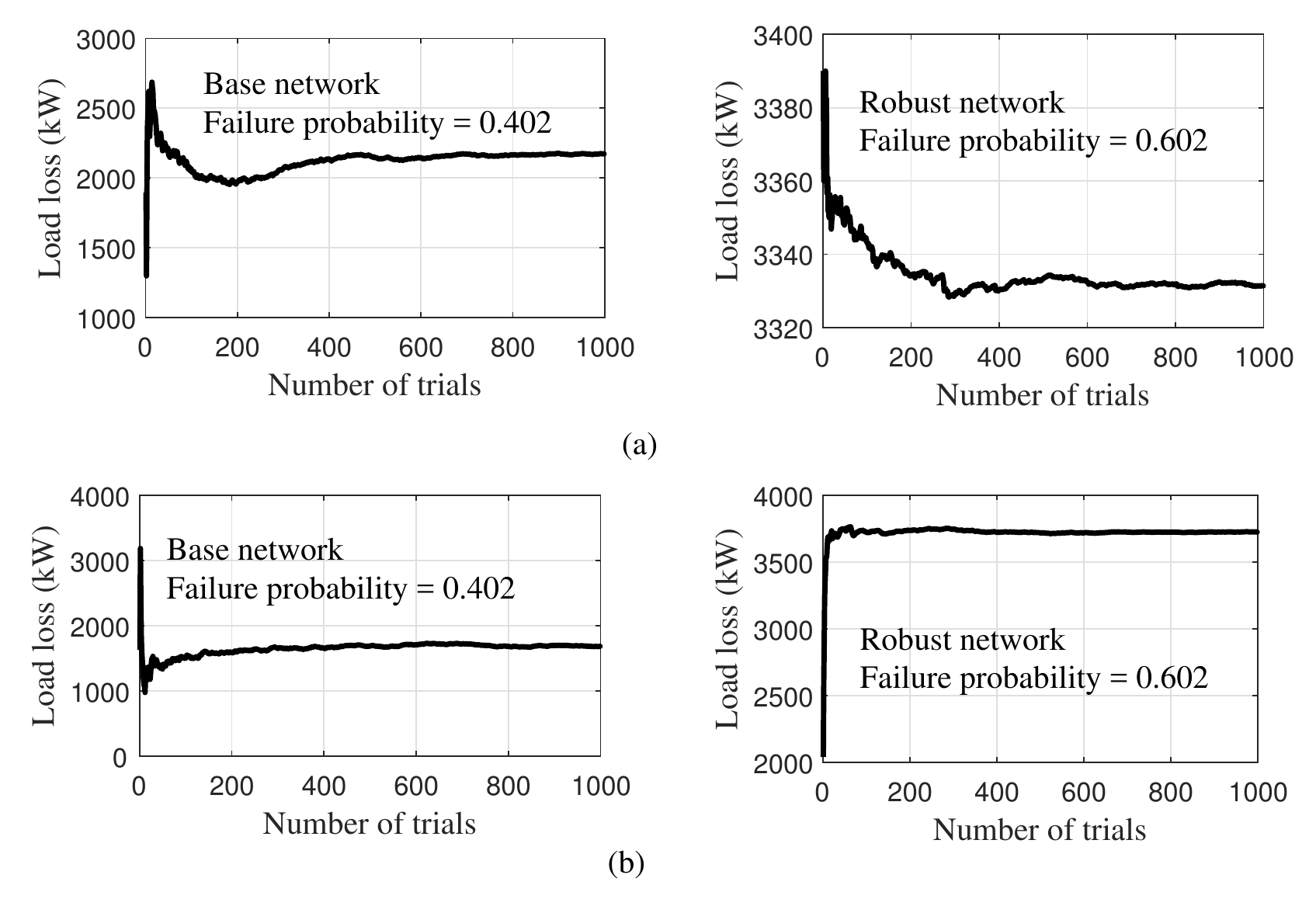}
    \vspace{-0.9 cm}
    \caption{Convergence of Monte-Carlo simulations for different scenarios in: (a) IEEE 123-bus and (b) IEEE 37-bus. It is observed that the convergence is guaranteed after 1000 trials in both test cases for different scenarios.}
    \label{fig:7}
    \vspace{-0.6 cm}
\end{figure}

\subsection{Parameters and Environment Setup}
 First, several Monte-Carlo simulations are done to obtain the required number of trials for the convergence of system loss function for a given damage scenario. It is observed from Fig. \ref{fig:7} that for both of the cases, 1000 trials are enough for the convergence. 
 
 The simulation steps are detailed here. First, we sample a wind speed from the PDF for wind speed profile (Fig. \ref{fig:2}) and obtain the failure probabilities for the distribution lines. We perform 1000 trials of Monte-Carlo simulations for the sampled wind speed. For each trial, a loss function is obtained by evaluating the phases of resilience curve as detailed in Section V. For a smart network with automated restoration, an optimization algorithm as detailed in \cite{8421055} is employed to assist with optimal island formation. Finally, the loss functions calculated for the 1000 Monte-Carlo runs are averaged. The process is repeated by sampling multiple wind speed values from the PDF for the wind speed profile. This helps model the system response for the different intensities of weather event (Fig. \ref{fig:2}). It is worth mentioning here that the duration of Phase-I is known and assumed to be 2 hours in this work, while the duration of Phase-II and Phase-III is provided by the simulation framework (see Section V). A risk-threshold of $\alpha = 0.95$ is selected for all simulations. 
 
\subsection{Evaluating Wind Speed Impacts on Operational Resilience}
\begin{figure}[t]
    \centering
    \includegraphics[width=0.5\textwidth]{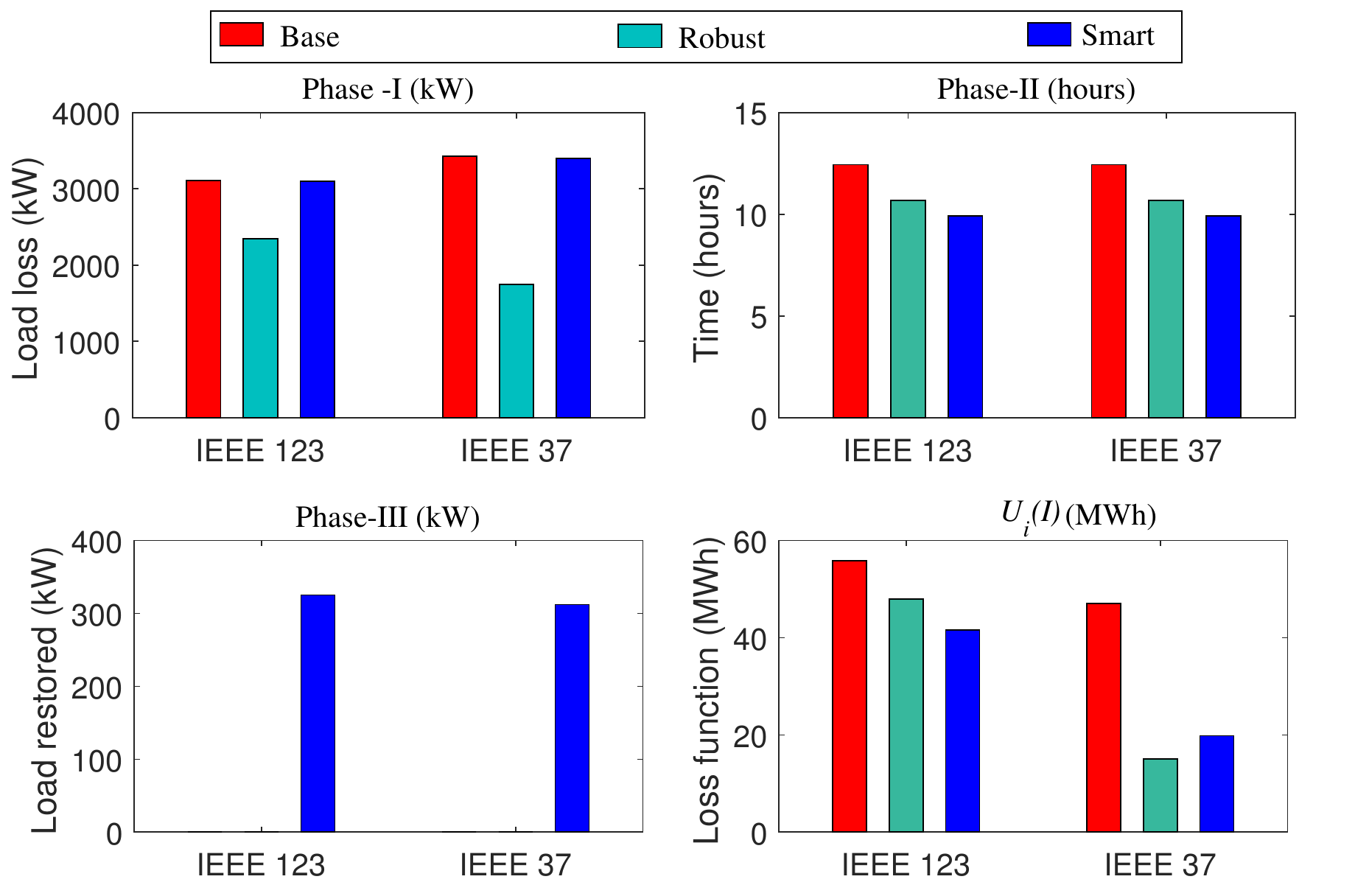}
    \vspace{-0.5 cm}
    \caption{Different phases resilience indicator for wind speed of 25 m/s }
    \label{fig:bar}
    \vspace{-1.0cm}
\end{figure}

In this section, we evaluate the resilience indicators in different phases of resilience curve for the three systems detailed in Section IV-B: base, robust and smart. A particular scenario of wind speed, $\omega$ = 25 m/s is considered for the simulations. Fig. \ref{fig:bar} shows the numeric values for three different phases of resilience curve under study and loss function based on the computed numeric values of each phase. It can be observed for both test cases that making a few distribution lines robust reduces the failure probability of lines during a high-speed wind event thus decreasing the load loss (kW) in Phase-I. Also, it is observed that making the network smarter by improving the response and enabling active islanding based restoration strategies results in a lower duration of post-event degraded state. For example, for a given damage scenario, the smart network takes 9.97 hours for damage assessment whereas the base network takes around 12.436 hours (See Phase-II of Fig. \ref{fig:bar}). Similarly, for the smart network, the restoration algorithm helps to restore loads using utility-owned DGs whereas the base network and robust network do not offer any restoration options in Phase-III of the resilience curve. A total of 325.13 kW and 312.22 kW load is restored in IEEE 123 and IEEE 37-bus systems respectively by suitable islanding scheme for the simulated scenario.

   \begin{table}[h]
        \centering
        \caption{VaR and CVaR for Different Weather Scenarios for IEEE 123-bus With $\alpha =95\%$}
        \label{table:1}
        \begin{tabular}{cccc}
            \toprule[0.4 mm]
            % \multicolumn{4}{c}{IEEE 123-bus}\\
            % \hline
            S.No&Weather Profile&$VaR_{\alpha}$ (MWh)&$CVaR_{\alpha}$(MWh)\\
            \hline
            1&Extreme &61.10&68.22 \\
            2&High &15.27&16.85 \\
            3&Normal&2.44&2.72 \\
            \toprule[0.4 mm]
        \end{tabular}
        \vspace{-0.5cm}
    \end{table}

 \subsection{Resilience Metric Computation: $VaR_\alpha$ and $CVaR_\alpha$}
 \subsubsection{Base Network} 
\begin{figure}[t]
    \centering
    \includegraphics[width=0.49\textwidth]{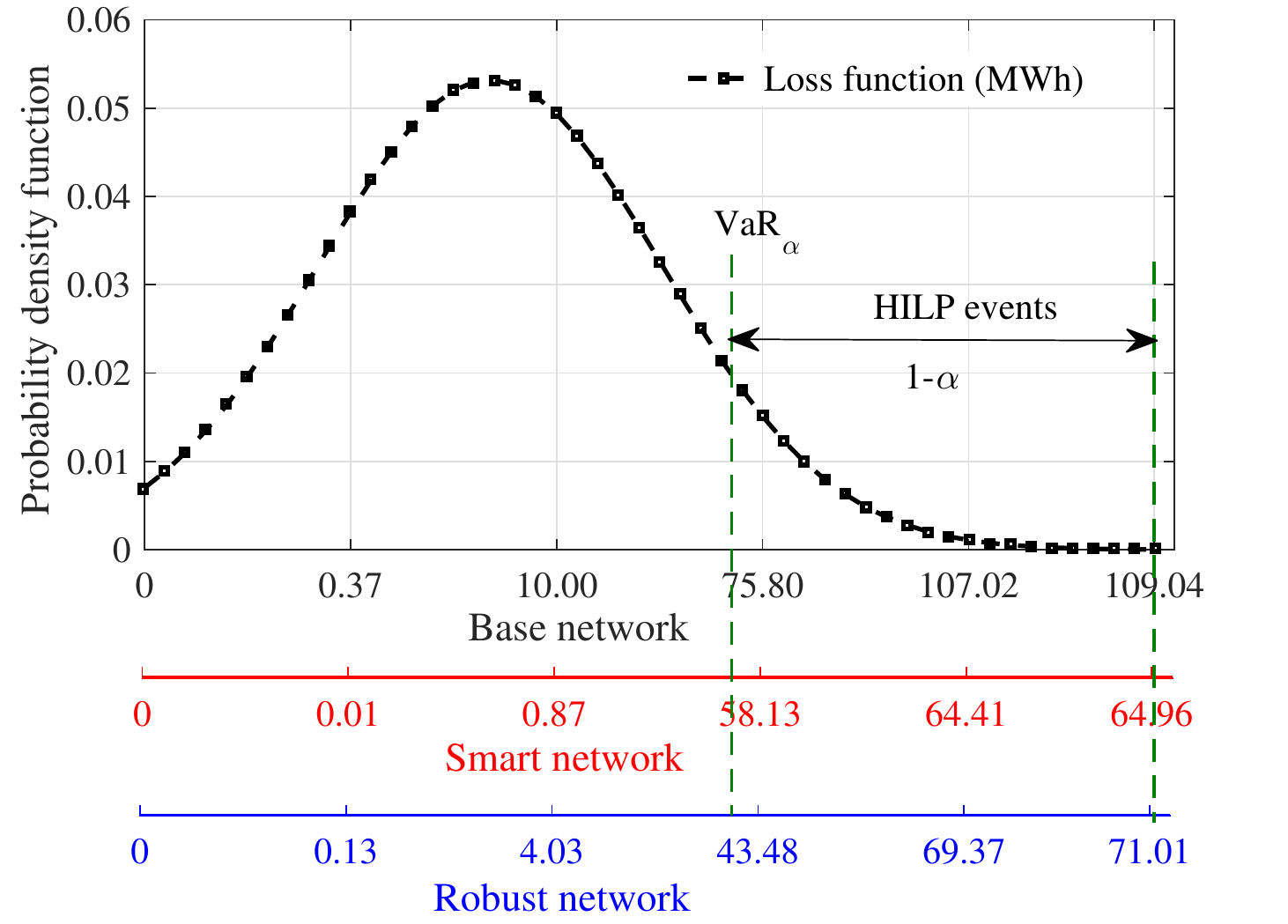}
    \vspace{-0.7cm}
    \caption{System performance loss function (MWh) for IEEE 37-bus test case during extreme wind profile for base, smart, and robust network.}
    \label{fig:37}
     \vspace{-0.2cm}
\end{figure}

\begin{figure}[t]
    \centering
    \includegraphics[width=0.49\textwidth]{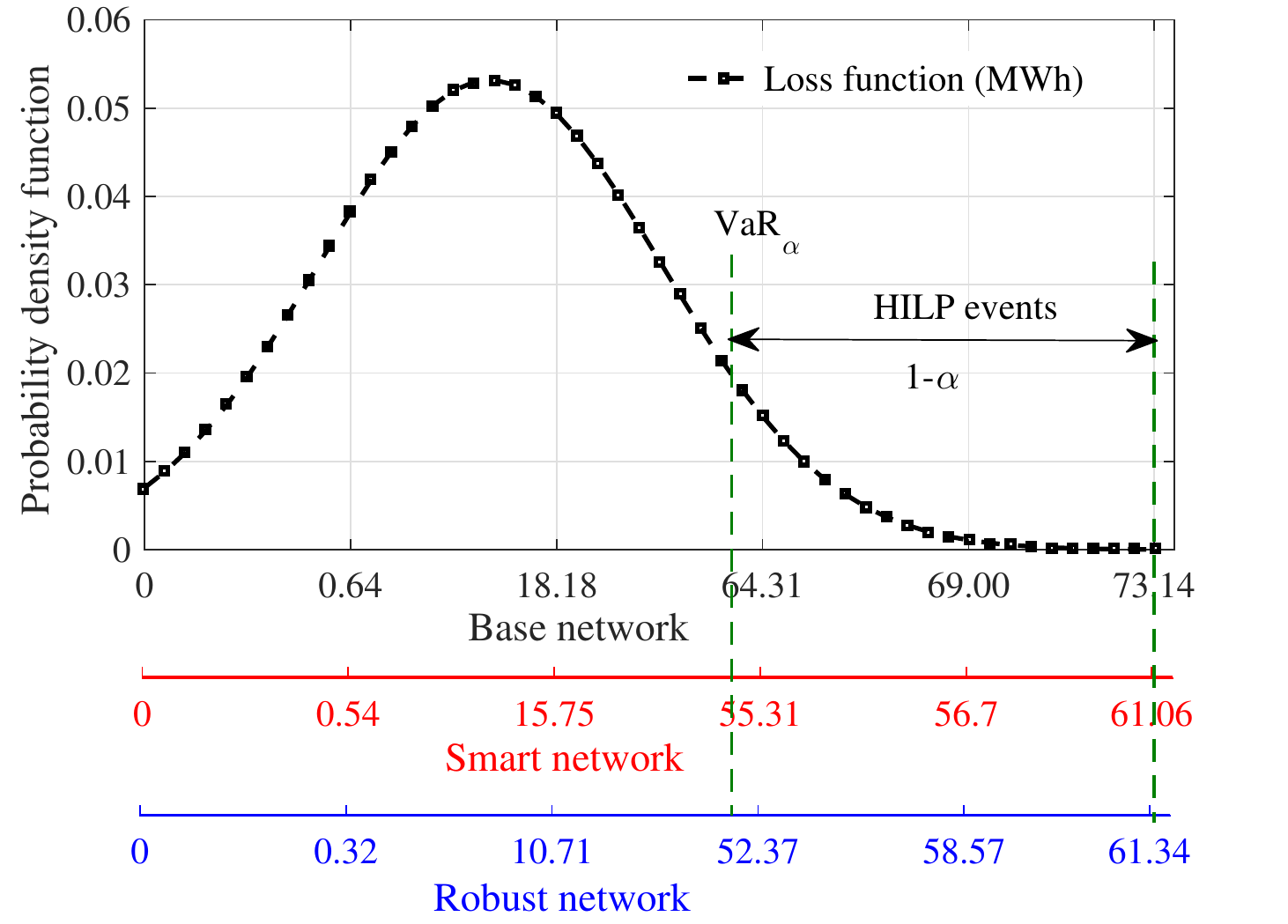}
    \vspace{-0.7cm}
    \caption{System performance loss function (MWh) for IEEE 123-bus test case during extreme wind profile for base, smart, and robust network.}
    \label{fig:123}
     \vspace{-1.cm}
\end{figure}

In this case, a base case network is assumed without any hardened lines or advanced restoration options using DGs. The loss function for each possible damage scenario in a particular region, i.e. a regional wind profile is obtained and plotted as a PDF (See black x-axes in Figs. \ref{fig:37} and \ref{fig:123}). The vertical dashed line in the figure represents the $VaR_\alpha$ which forms a baseline for evaluation of expected resilience loss in HILP event. Table \ref{table:1} reports the $VaR_\alpha$ and $CVaR_\alpha$ for the three wind profiles in Fig. 2. It is observed that the expected resilience loss (MWh) beyond a threshold, characterized by $CVaR_\alpha$, is higher for an extreme wind profile and lower for normal wind profile. This illustrates that the proposed approach can effectively quantify the resilience for weather events specific to a given geographical area. Note that while evaluating the resilience metric for the test cases, we only consider extreme wind profile.

 \subsubsection{Proactive Planning Measures} In this case, two different planning measures are implemented for both test feeder and resilience improvement is analyzed for the systems when impacted by the wind event corresponding to extreme weather profile. 
 First, the distribution networks under study are assumed to have improved response and smart restoration actions with DGs and tie switches (see Fig. \ref{fig:6}). That is, for any damage scenario, the system recovers with smart restoration actions using algorithm detailed in \cite{8421055}. This planning measure helps to enhance resilience by modifying Phase-II and Phase-III of the resilience curve. The loss function for each wind speed is calculated using Monte-Carlo simulations and mapped into the PDF of wind intensity. The system loss is plotted as x-axis (in red) in Fig. \ref{fig:37} and Fig. \ref{fig:123}.  
 
 In the second case, the distribution network is made robust via reinforcing/hardening a few distribution lines as previously discussed. For this case, the loss function is computed and mapped into PDF as shown in the blue x-axis of Fig. \ref{fig:37} and Fig. \ref{fig:123}. Notice that in both cases, the black x-axis represents the probabilistic loss function for the base case. 

 Once the PDF for system loss is generated, $\alpha=0.95$ is used to capture the HILP events or the extreme cases of regional wind profile. $VaR_{\alpha}$ and $CVaR_{\alpha}$ are calculated as shown in Table \ref{table:2}. Note that the numerical values in the table report the maximum ($VaR_{\alpha}$) and expected loss ($CVaR_{\alpha}$) of energy loss (in MWh) due to $(1-\alpha)\%$ of highest impact events. It can be observed that for both test cases, $VaR_{\alpha}$ and $CVaR_{\alpha}$ values are reduced for the network with an improved restoration plan and the network with added reinforcement measures. The proposed metric, therefore, quantifies the operational resilience of a power distribution network and can be used as an effective tool for comparing different strategies for boosting grid resilience.

    \begin{table}[t]
        \centering
        \caption{VaR and CVaR For Different Cases With $\alpha =95\%$}
        \label{table:2}
        \begin{tabular}{cccc}
         \toprule[0.3 mm]
         \rowcolor{lightgray}\multicolumn{4}{c}{Test case: IEEE 123-bus}\\
         \toprule[0.3 mm]
            S.No&Network&$VaR_{\alpha}$(MWh)&$CVaR_{\alpha}$(MWh)\\
            \hline
            1&Base Network &61.10&68.22 \\
            
            2&Smart Network &52.53&56.73 \\
            
            3&Robust Network & 50.25&57.52 \\
            \toprule[0.3 mm]
            \rowcolor{lightgray}\multicolumn{4}{c}{Test case: IEEE 37-bus}\\
             \toprule[0.3 mm]
             S.No&Network&$VaR_{\alpha}$(MWh)&$CVaR_{\alpha}$(MWh)\\
            \hline
            1&Base Network &67.89&101.09 \\
           
            2&Smart Network &44.63&63.23 \\
            
            3&Robust Network &34.97&64.61\\
            \toprule[0.3 mm]
        \end{tabular}
        \vspace{-0.9cm}
    \end{table}
    
\begin{table}[h]
        \centering
        \caption{VaR and CVaR for IEEE 123-bus with Different DG Location}
        \label{table:3}
        \begin{tabular}{c|cc|cc}
         \toprule[0.3 mm]
         \hline
          \multirow{2}{*}{Network}&\multicolumn{2}{c|}{DG-\{250, 450, 95\}}&\multicolumn{2}{c}{DG-\{250, 450, 48\}}\\
           \cline{2-5}
            &$VaR_{\alpha}$&$CVaR_{\alpha}$&$VaR_{\alpha}$&$CVaR_{\alpha}$\\
            \hline
            Smart Network &52.53&56.73&41.39&52.13 \\
            \toprule[0.3 mm]
        \end{tabular}
        \vspace{-0.9cm}
    \end{table}   

\subsection{Discussions on Effect of DG Location} 
In this work, we have randomly placed DGs in the distribution grid to evaluate the proposed framework in computing the proposed resilience metrics. To study the effect of DG location, we have changed the location of one of the DGs in IEEE 123-bus network from bus 95 to bus 48. The results are summarized in Table \ref{table:3}. It is observed that the change in DG location results in a new value for resilience metric. This is because, with the change in DG location, a different DG supplied island is formed to pick loads during the restoration in the aftermath of an event. Therefore, the location of DG is crucial towards the overall resilience of the system and should be optimized.

The primary objective of this work is to propose risk-based metrics and to demonstrate their use in quantifying system resilience during HILP events and under different resource allocations for resilience improvement. Therefore, in this work, we do not optimize the DG location or size but simply demonstrate how the metric can be used to evaluate the operational resilience for any possible system configuration by randomly locating and sizing the DGs. The problem of optimal resource allocation to improve resilience is an important direction for future research.

\vspace{-0.3cm}
\subsection{Discussions on Proposed Metric and Planning Measures} 
\begin{figure}[h]
    \centering
    \includegraphics[width=0.48\textwidth]{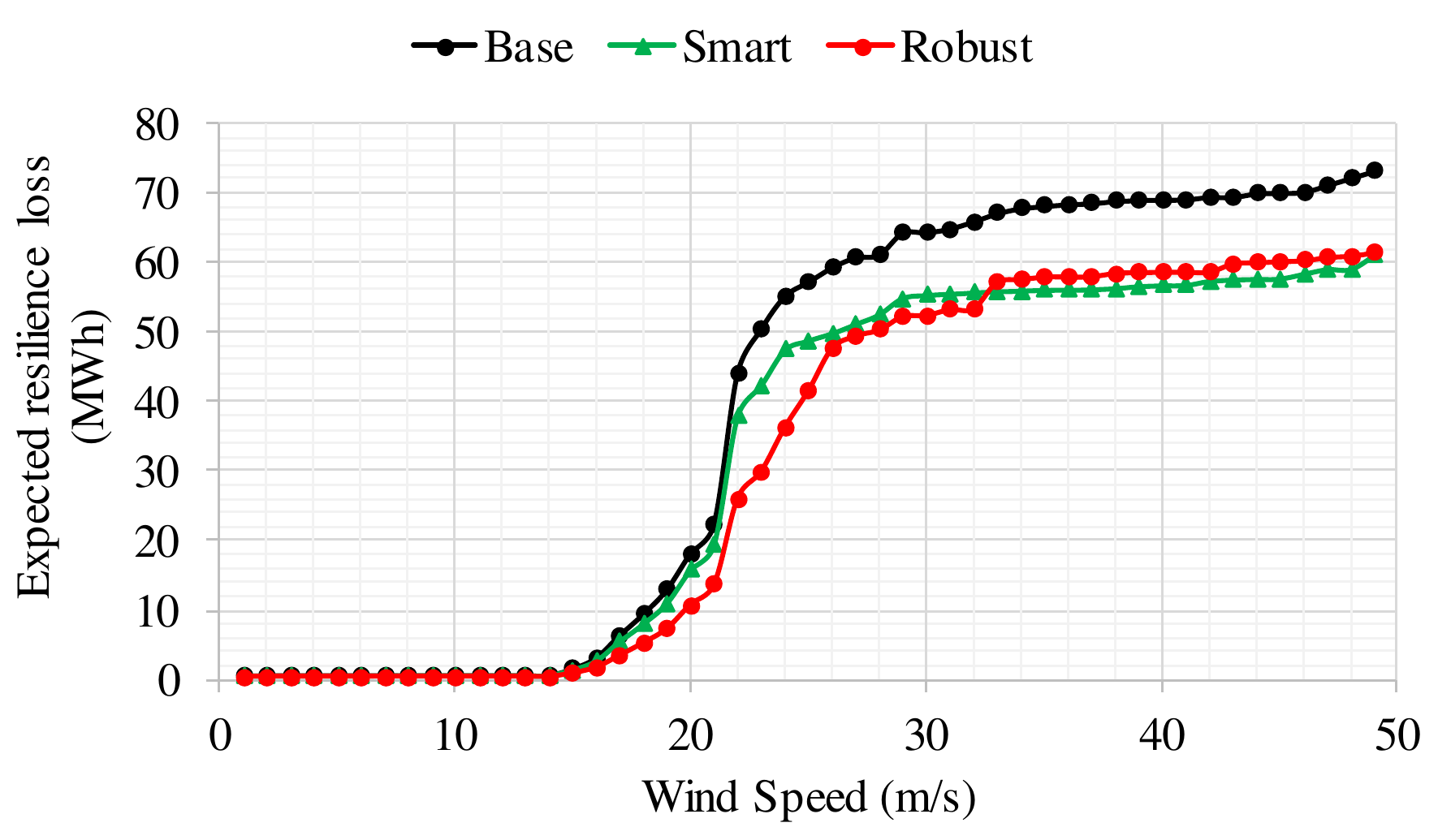}
    \vspace{-0.3cm}
    \caption{Expected loss in resilience for different wind speed for IEEE 123-bus test case}
    \label{fig:10}
     \vspace{-0.0cm}
\end{figure}

The resources allocated for resilience enhancement can either lessen the impacts of an extreme event or improve the system recovery time. Minimizing the expected resilience loss requires an optimal allocation of planning resources that help simultaneously reduce both the system loses and recovery time. Although in this paper, we do not assess the associated costs of the investment decisions (smart vs. robust network), it is well understood that hardening measures are significantly more expensive than smart grid investments. However, smart operational measures alone are not enough to keep the lights on during an extreme event \cite{panteli2015grid}. Furthermore, it can be observed in Fig. \ref{fig:10} that for different event intensities, the optimal system response alternates between implementing smart actions and employing infrastructure hardening solutions. Since weather events and their component-level impacts are rather stochastic, having an optimal planning measure including both smart actions and hardening can provide the most suitable road-map for improving the resilience. 

The proposed metrics ($CVaR$ and $VaR$) can be employed to optimally allocate resources for different planning activities to enhance the grid's operational resilience. Specifically, the proposed metric will be used to plan for the resources that will improve the system performance during HILP events. This requires a mechanism to quantify the improvements in risks associated with the highest impact events under a given resource allocation. The proposed resilience metrics specifically model and quantify these aspects making them suitable for resource planning for resilience. Towards this goal, we envision a risk-averse framework for resource planning to manage disruptions in the power distribution grid and to inform on the optimal planning decisions.

Specifically, the planning problem for resilience can be modeled as conditional value-at-risk (CVaR) optimization problem, a method widely used for risk-averse financial planning \cite{ref_Rockafellar,ref_Krokhmal,ref_Uryasev} which is apt when HILP events are a primary concern. The main idea is to optimize a $CVaR$ metric, which measures the risks associated with loss of resilience when the system under study is subjected to a percentage of highest-impact events. This is achieved by developing functional dependence between the event's impacts and system loss and recovery functions under specified resource allocations. For example, the decision from such planning problem can help identify the potential locations of distributed generators and potential lines to upgrade that will minimize the disruptions to system's critical loads during a future extreme event. Since the planning is based on the system's response to HILP events, the resulting upgraded grid will ensure that the supply to critical loads is more resistant to disasters.

\section{Conclusions}
This paper presents probabilistic metrics and a detailed simulation approach to quantify the resilience of electric power distribution systems. The expected resilience loss due to HILP events is characterized using two risk-based metrics: $VaR_\alpha$ and $CVaR_\alpha$. By definition, these metrics quantify the worst outcomes of low probability highest impact events such as an extreme weather event making them suitable for measuring operation resilience of power distribution grid. We also detail an approach to quantify the effects of smart operational measures and infrastructure hardening solutions on improving the distribution system resilience. From simulation studies, it is observed that the proposed resilience metrics are well-suited to compare different resilience enhancement strategies.
Finally, we conclude with an observation that a hybrid network (stronger and smarter) might offer better resilience while optimizing the investment costs for the possible resilience enhancement measures.  

A potential direction for future research is the use of proposed operational resilience metrics in optimally allocating planning resources to improve the system's performance during HILP events. Specifically, a risk-averse framework is envisioned for resource planning to manage disruptions in the power distribution grid, and to inform on the optimal planning decisions.

\bibliographystyle{IEEEtran}
\bibliography{references}
\vspace{-1.0 cm}

\begin{IEEEbiography}[{\includegraphics[width=1in,height=1.25in,clip,keepaspectratio]{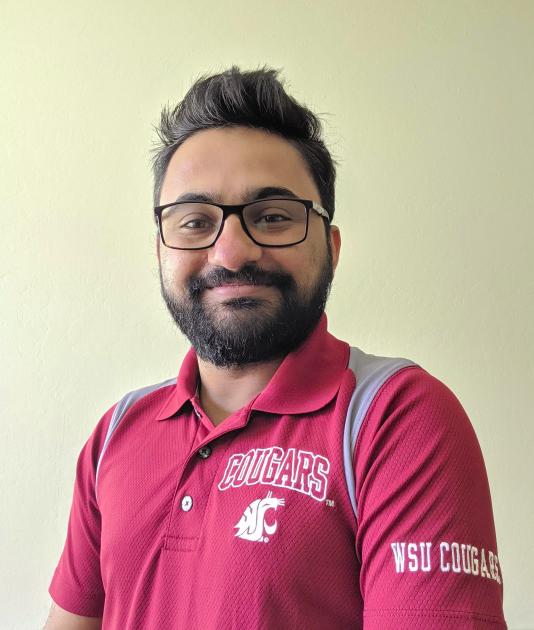}}]{\textbf{Shiva Poudel}} (S'15) received the B.E. degree from the Department of Electrical Engineering, Pulchowk Campus, Kathmandu, Nepal, in 2013, and the M.S. degree from the Electrical Engineering and Computer Science Department, South Dakota State University, Brookings, SD, USA, in 2016. He is now pursuing the Ph.D. degree in the School of Electrical Engineering and Computer Science, Washington State University, Pullman, WA. 
In 2018 and 2019, he was a summer intern with Mitsubishi Electric Research Laboratories, Cambridge, MA, USA and Electric Power Research Institute, Palo Alto, CA, USA respectively.
His current research interests include distribution system restoration, resilience assessment, and distributed algorithms.
\end{IEEEbiography}

\vspace{-0.5 cm}

\begin{IEEEbiography}[{\includegraphics[width=1in,height=1.25in,clip,keepaspectratio]{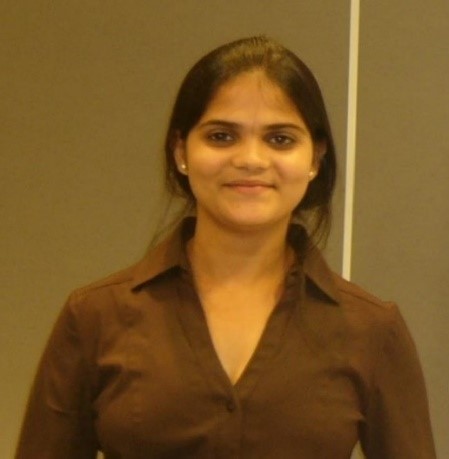}}]{\textbf{Anamika Dubey}} (M'16) received the M.S.E and Ph.D. degrees in Electrical and Computer Engineering from the University of Texas at Austin in 2012 and 2015, respectively. Currently, she is an Assistant Professor in the School of Electrical Engineering and Computer Science at Washington State University, Pullman.

Her research focus is on the analysis, operation, and planning of the modern power distribution systems for enhanced service quality and grid resilience. At WSU, her lab focuses on developing new planning and operational tools for the current and future power distribution systems that help in effective integration of distributed energy resources and responsive loads. 
\end{IEEEbiography}

\vspace{-0.5 cm}
\begin{IEEEbiography}[{\includegraphics[width=1in,height=1.25in,clip,keepaspectratio]{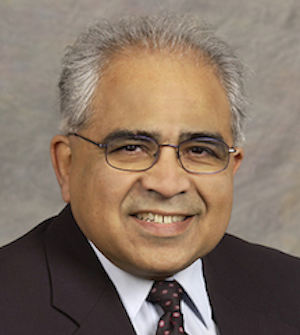}}]{\textbf{Anjan Bose}} (LF'89) received the B. Tech. (Hons.) degree from the Indian Institute of Technology,
Kharagpur, Kharagpur, India; the M.S. degree from the University of California, Berkeley, CA, USA;
and the Ph.D. degree from Iowa State University, Ames, IA, USA.
He was with the industry, academia, and government for 40 years in power-system planning, operation, and control. He is currently a Regents Professor and an endowed Distinguished Professor in Power
Engineering with Washington State University, Pullman, WA, USA. Dr. Bose is the recipient of the Herman Halperin Award and the Millennium Medal from the IEEE. He is a member of the National Academy of Engineering.
\end{IEEEbiography}

\end{document}